# Magnetic light amplification by stimulated emission of radiation in subwavelength systems of a dielectric cavity and magnetic quantum emitters

Zhong-Jian Yang*, Xiao-Jing Du, Ma-Long Hu, and Jun He*

*Hunan Key Laboratory of Nanophotonics and Devices, School of Physics and Electronics, Central South University, Changsha 410083, China*

*E-mail: zjyang@csu.edu.cn; junhe@csu.edu.cn

**Abstract:** We propose a magnetic laser in a subwavelength system consisting of a high-refractive-index dielectric cavity and an active medium formed by magnetic quantum emitters. Stimulated emissions of magnetic quantum emitters induced by their coherent interactions with quantized magnetic fields of a cavity are theoretically considered. The condition to achieve such a magnetic laser is obtained. Numerical results show that magnetic lasers are feasible in some realistic systems, for example, a silicon disk of high-quality whispering gallery modes with embedded emitters. Furthermore, the competitions between the electric interaction and magnetic one in terms of their Purcell factors are also considered in some magnetic laser achievable systems. In a wavelength-scale silicon block of a high-order magnetic mode, the ratio of magnetic Purcell factor to the electric one can reach more than $\sim 10^3$ large. Our results open up ways to enhanced magnetic light-matter interactions.

## I. INTRODUCTION

The interactions of the magnetic component of light and magnetic dipole (MD) transitions at optical frequencies are usually several orders of lower than their electric counterparts [1, 2]. Hence, the light-matter interactions are generally interpreted as the couplings of electric fields and electric dipoles. Nevertheless, the magnetic interactions provide another dimension in glimpsing light-matter interactions [2-4]. Furthermore, strong MD transitions of quantum emitters at optical frequencies are indeed found in some lanthanide series ions such as $Eu^{3+}$ and $Er^{3+}$ [5-7]. The interactions of these magnetic quantum emitters (MQEs) with light have been attracting increasing research interests despite high technology requirements [4, 8-12].

Putting MQEs close to photonic structures could allow one to largely turn the couplings between MQEs and light. The spontaneous decay rate enhancement, which is also termed as Purcell factor [13-15], can be modified a lot [4]. Due to their high near-field confinement, plasmonic nanostructures have been utilized to interact with MQEs [4, 16-18]. However, plasmonic systems suffer from high material losses, and complex geometries are required to obtain effective magnetic responses. In the past years, all-dielectric (sub)wavelength-scale structures with high-refractive indexes $n_r$ have been found to exhibit Mie-like resonances [19, 20], and they can readily support magnetic near-field responses. Those magnetic responses provide a platform for magnetic light-matter interactions [21-24]. The common resonant modes in the reported subwavelength all-dielectric structures are low-order electromagnetic multipoles such as electric/magnetic dipoles, toroidal modes, and supercavity mode

[19, 20, 25-28]. Most of these modes show low quality ($Q$) factors ($\sim 10^1$) and low electromagnetic near field enhancement ($\sim 10^1$) with a common material silicon (Si) $n_r \sim 3.5$. Recently, it has been demonstrated that subwavelength dielectric resonators ($n_r \sim 3.5$) can also support whispering gallery modes (WGMs) with high enough $Q$ factors ($\sim 10^5$) and high electromagnetic near field enhancements ($\sim 10^2$) [29]. These achievements of all-dielectric magnetic cavities hold great promise to enable more efficient magnetic photons-MQEs couplings [30, 31].

Here, we theoretically propose that a magnetic laser can be obtained in a subwavelength system of a dielectric cavity and MQEs. MQEs are modeled as simple two-level emitters with MD transitions. The MQEs can undergo stimulated emission of radiation though coherent interactions with the photons in the cavity. The stimulated emission is similar to that in common lasers or spasers [32-40] while the couplings here are magnetic interactions instead of the electric ones. A subwavelength dielectric disk with high-$Q$ WGMs is numerically considered. The magnetic laser can be obtained due to the high-$Q$ features of the WGMs. Furthermore, we also consider the competition between the magnetic interactions and electric ones in some magnetic laser achievable systems. Specifically, the ratio of magnetic Purcell factor to the electric one in a subwavelength silicon block of a high-$Q$ magnetic mode can reach $\sim 10^3$ large. This property makes such a kind of dielectric cavity a suitable platform to carry out the magnetic light-matter interactions including a magnetic laser.

## II. THEORETICAL APPROACH

The magnetic field of a high-$Q$ resonant mode of a dielectric cavity can be quantized based on that of the standard harmonic oscillators [37, 41, 42] (see Appendix A). A MQE is taken as a two-level emitter with a matrix element of $\vec{M}_{10}$ for its MD transition. Then, the interaction Hamiltonian between a MQE and the quantized magnetic field under the rotating-wave approximation can be expressed as

$$H_{int} = \hbar g(\hat{a}\hat{\sigma}_{10}e^{-i(\omega_n t+\varphi(r))} + \hat{a}^+\hat{\sigma}_{01}e^{i(\omega_n t+\varphi(r))}), \tag{1}$$

where $\omega_n$ is the frequency of the photon, and $\hat{a}^+$ and $\hat{a}$ are the creation and annihilation operators of a photon, respectively. $\hat{\sigma}_{10}$ and $\hat{\sigma}_{01}$ are transition operators of the MQE. $\varphi(r)$ represents the spatial phase of the magnetic field at the location $r$. $g$ is the coupling strength $g = \sqrt{\frac{\mu_0 \omega_n}{2\hbar V_m}}\vec{M}_{10} \cdot \frac{\vec{B}(r)}{B_{max}}$ (see Appendix A), where $V_m$ is the magnetic field mode volume of the cavity $V_m = \frac{\int \mu_0 |\vec{H}(r)|^2 d^3 r}{\mu_0 H_{max}^2}$, $\vec{B}(r)/B_{max}$ $(= \vec{H}(r)/H_{max})$ is the normalized magnitude of the magnetic field in the cavity, $\vec{H}(r)$ is the magnitude of the magnetic field at location $r$, $\hbar$ is reduced Planck constant, and $\mu_0$ is the permeability of vacuum. The analytical description is similar to that of electric interactions [38, 43-46], while the coupling strength $g$ should be replaced by $g_e = \sqrt{\frac{\omega_n}{2\hbar\varepsilon_0 V_e}}\vec{\mu}_{10} \cdot \frac{\vec{E}(r)}{E_{max}}$ for an electric interaction. $\vec{\mu}_{10}$ is the matrix element of the electric dipole transition of an electric quantum emitter (EQE), $\vec{E}(r)/E_{max}$ is the normalized electric field, $\varepsilon_0$ is the permittivity of vacuum, and $V_e$ is the electric-field mode volume of a cavity mode.

Under Fermi's golden rule, the total emission rate into photons from a MQE can be expressed as (see Appendix B)

$$\Gamma' = 2\pi g^2 (N_n + 1) \int F(\omega)\frac{\gamma_n^2}{(\omega-\omega_n)^2+\gamma_n^2}d\omega, \tag{2}$$

where the term $N_n$+1 represents the contributions from the stimulated $\Gamma^{st}$ ($N_n$) and spontaneous $\Gamma^{sp}$ (1) emissions. $\int F(\omega) \frac{\gamma_n^2}{(\omega-\omega_n)^2+\gamma_n^2} d\omega$ is the spectral overlap factor [37], where $F(\omega)$ is the normalized-to-1 spectrum of MD transitions, and $\gamma_n$ is the relaxation rate of the photon. $F(\omega)$ is highly related to the relaxation rate ($\gamma_{10}$) of the MQE. For $\gamma_{10} \ll \gamma_n$, the overlap factor is 1. While for $\gamma_{10} \gg \gamma_n$, the overlap factor becomes $\gamma_n/\gamma_{10}$. Note that in the above integrations of overlap factors, we have assumed the resonant couplings between the MQE and photons. The stimulated absorption rate $\Gamma^{sa}$ is equal to the stimulated emission rate $\Gamma^{sa} = \Gamma^{st}$. For the couplings of many MQEs and photons, the generation rate of photon number $N_n$ can be expressed as

$$\dot{N}_n = \int \Gamma^{st} \rho_{eff}(r) d^3r + \int \Gamma^{sp} \rho_1(r) d^3r - N_n \gamma_n, \tag{3}$$

where $\rho_{eff}(r) = \rho_1(r) - \rho_0(r)$, $\rho_1(r)$ and $\rho_0(r)$ are the population densities of MQEs in the excited and ground states, respectively. The rate equation for the population of MQEs can be written as (see Appendix C)

$$\int \dot{\rho}_{eff}(r) d^3r =$$

$$\int (W_{01} - \Gamma^{sp})(\rho_1(r) + \rho_0(r)) d^3r - \int (W_{01} + \Gamma^{sp} + 2\Gamma^{st}) \rho_{eff}(r) d^3r, \tag{4}$$

where $W_{01}$ is the pumping rate of the MQEs.

The coupling strength $g$ is an important parameter that determines if a magnetic laser can be realized in a system. Generally, $g$ can be obtained by calculating the mode volume based on numerical methods, for example, the finite-difference time-domain (FDTD) simulations. Alternatively, the spontaneous decay rate enhancement can be simulated by numerical methods. Then, $g$ can also be obtained correspondingly. In the FDTD simulations, the calculations are carried out with the condition of $\gamma_{10} \ll$

$\gamma_n$, and the directly simulated decay rate enhancement is $2\pi g^2/\Gamma_1$, where $\Gamma_1$ is the decay rate of a MQE in the medium of $n_r$ [31]. $\Gamma_1 = n_r^3\Gamma_0$, where $\Gamma_0$ is the vacuum spontaneous decay rate of a MQE [4]. Thus, the magnetic Purcell factor is $\Gamma^{sp}/\Gamma_0 = 2\pi g^2/\Gamma_0 = 2\pi g^2 n_r^3/\Gamma_1$.

In our considered systems, a photon is generated through stimulated or spontaneous emission from MQEs. Meanwhile, a photon can be annihilated through stimulated absorption by MQEs or its own relaxation in the cavity (Eq. (3)). The net generation rate of coherent photons in a system is $\int \rho_{eff}(r)\Gamma^{st}d^3r$. The magnetic laser can occur when this value is larger than the photon relaxation rate $N_n\gamma_n$, namely,

$$\int \rho_{eff}(r)\Gamma^{st}d^3r > N_n\gamma_n. \qquad (5)$$

If we assume that the coupling strength $g$ of each MQE and cavity is the same for simplicity. Under the situation of $\gamma_{10} \ll \gamma_n$, the net generation rate of coherent photons becomes $N_{eff}\Gamma^{st}$. Here, $N_{eff}$ is the inversed total number of MQEs $N_{eff} = \int \rho_{eff}(r)d^3r$. Eq. (5) then reduces to

$$N_{eff}2\pi g^2 > \gamma_n. \qquad (6)$$

Thus, $\gamma_n/2\pi g^2$ represents the threshold number of MQEs required to achieve a magnetic laser in the above considered system. Under the situation of $\gamma_{10} \gg \gamma_n$, the net generation rate of coherent photons is $N_{eff}\Gamma^{st} \gamma_n / \gamma_{10}$. The condition to realize a magnetic laser becomes $N_{eff}2\pi g^2 > \gamma_{10}$ correspondingly. For the rest of discussion, we will take the situation of $\gamma_{10} \ll \gamma_n$ unless specified.

### III. MAGNETIC LASER IN A SUBWAVELENGTH WGM CAVITY

A realistic dielectric cavity with active MQEs is considered as shown in Fig. 1(a).

The cavity is a Si disk supporting high-$Q$ subwavelength WGMs [29]. The radius and the height are both 630 nm. The geometry is chosen to match the wavelength region such that the refractive index of Si is around $n_r$ = 3.5. Simulations show that the resonance of a TE WGM of the azimuthal mode index $m$ = 7 occurs at $\lambda$ = 1230 nm. The $Q$-factor of this mode is 1.5 x $10^5$ and the corresponding relaxation rate is $\gamma_n$ = 1.6 x $10^9$ s$^{-1}$. The magnetic Purcell factor $2\pi g^2/\Gamma_0$ can reach 1.65 x $10^5$. Note that the diameter of the above disk is very close to the resonant wavelength of $m$ = 7 mode, while the height is much smaller than this wavelength. Thus, one can roughly take this mode as a subwavelength resonance. In fact, by slightly increasing the height and decreasing the diameter, a WGM mode will keep the same resonant wavelength and almost the same $Q$-factor, while both the diameter and height are completely subwavelength [29]. In this work, the height and radius are taken the same for simplicity.

We assume the maximum population inversion of active MQEs $\rho_1 \gg \rho_0$. Thus, $N_{eff} \approx N_{MQE} = \int(\rho_1(r) + \rho_0(r))d^3r$, where $N_{MQE}$ represents the total number of MQEs. The coupling strength $g$ of each MQE and cavity is assumed to be the same for simplicity. With the situation of low enough temperature ($\gamma_{10} \ll \gamma_n$) and a realistic value of $\Gamma_0$ = $10^1$ s$^{-1}$ [7], the threshold number of MQEs required to achieve a magnetic laser is $N_{MQE}^{th} \approx \gamma_n/2\pi g^2 \approx 9.8$ x $10^2$ based on Eq. (6). Such a threshold value should be easily satisfied experimentally. The $N_{MQE}^{th}$ for a magnetic laser increases dramatically as the $Q$-factor of a cavity mode decreases ($N_{MQE}^{th} \sim 1/Q^2$, Fig. 1(b)), and reaches more than ~$10^8$ for $m$ = 3 ($Q \approx 150$). For the situation of $\gamma_{10} \gg \gamma_n$, the $N_{MQE}^{th}$ becomes $N_{MQE}^{th} = \gamma_{10}/2\pi g^2$. This value is relatively $\gamma_{10}/\gamma_n$ times larger than that under the situation of low enough temperature.

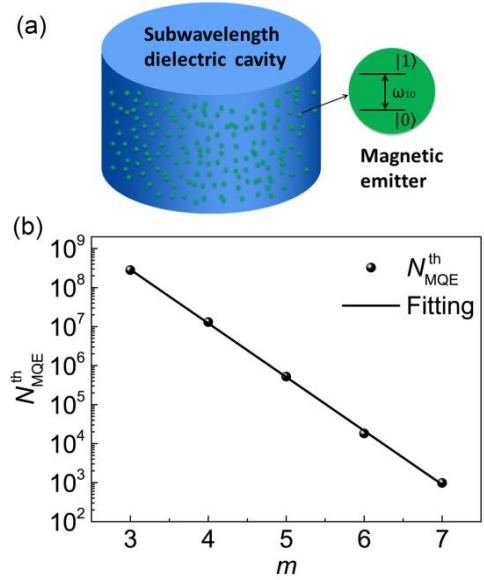

**Fig. 1**. (a) Schematic of a magnetic laser system consisting of a subwavelength dielectric cavity and active MQEs. Each MQE is a two-level emitter. (b) $N_{MQE}^{th}$ of a WGM-resonant Si cavity with different mode *m*. Each WGM is a TE mode. The line is the fitting results with an exponential decay function. The radius and the height are both 630 nm.

By enlarging the size of a disk cavity, it becomes relatively easier to obtain a magnetic laser. The WGM response is a geometric resonance in a dielectric cavity. Thus, the resonant wavelength $\lambda_n$ of a WGM is in proportion to the disk diameter *D* ($\lambda_n \propto D$) [29], where the height/diameter ratio of a disk is always kept the same. The *Q*-factor of each WGM remains the same with varying the disk size (Fig. 2(a)). The magnetic field distribution size increases proportionally with the disk size. Thus, the mode volume $V_m$ increases proportionally with the geometric volume of the disk ($V_m \propto D^3$). By combining all these factors, the Purcell factor $\frac{\Gamma^{sp}}{\Gamma_0} = \frac{2\pi g^2}{\Gamma_0} \propto Q\lambda_n^3/V_m$ [4] also remains unchanged with disk size. This is also confirmed by direct simulations (Fig. 2(a)). The decay rate of a WGM photon $\gamma_n$ decreases with the disk size $\gamma_n \propto 1/D$ as the resonant frequency $\omega_n$ decreases with $D$ ($\omega_n \propto 1/D$) while the $Q$-factor remains unchanged. Thus, the $N_{MQE}^{th}$ to achieve a magnetic laser becomes relatively smaller with disk size $N_{MQE}^{th} = \gamma_n/2\pi g^2 \propto 1/D$ (Fig. 2(b)). Here, we have assumed a fixed $\Gamma_0$ for simplicity. The above analysis is not restricted to a specific WGM. Thus, the relation $N_{MQE}^{th} \propto 1/D$ with a fixed $\Gamma_0$ applies for any WGM in the disk system. Furthermore, a larger disk also provides more space to host the MQEs. This is also an important profitable factor to realize such a system in experiments.

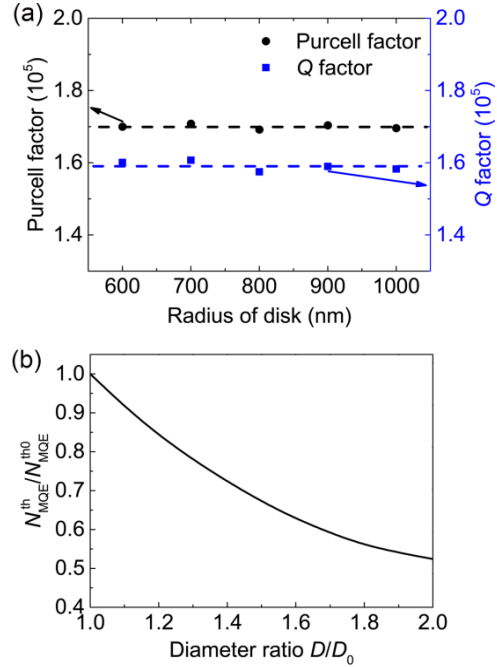

**Fig. 2**. (a) Directly simulated $Q$ factor (blue dots) and magnetic Purcell factor $\Gamma^{sp}/\Gamma_0$ (black dots) as a function of the disk size (radius). The WGM is a $m = 7$ TE mode for each case. (b) Normalized threshold number of MQEs $N_{MQE}^{th}/N_{MQE}^{th0}$ as a function of the size (diameter) ratio $D/D_0$ (red line). $D_0$ is a reference diameter of a disk with a threshold number of $N_{MQE}^{th0}$. The height is always kept the same the radius in each case.

Now let us turn to the number of photons $N_n$ of the magnetic laser system. At first, $N_n$ increases as the gain is larger than the loss. Then, $N_n$ turns to be saturated (denoted by $N_n^{max}$) when the gain equals the loss. The $N_n^{max}$ can be obtained by solving Eqs. (3) and (4) under the steady-state conditions, namely, $\dot{\rho}_{eff} = 0$ and $\dot{N}_n = 0$. The analytical expression of $N_n^{max}$ as a function of the system parameters is complex (see Appendix D). There are two solutions for $N_n^{max}$, where the negative one is omitted. The $N_n^{max}$ as a function of pumping rate for cases with different number of MQEs $N_{MQE}$ are shown in Fig. 3. When $N_{MQE}$ is more than several times larger than the $N_{MQE}^{th}$, $N_n^{max}$ increases linearly with the pumping rate $W_{01}$ ($N_n^{max} \propto (N_{MQE} - N_{MQE}^{th})W_{01}/\Gamma^{sp}$). For the case where $N_{MQE}$ is equal to the $N_{MQE}^{th}$, $N_n^{max}$ shows a square root function of the pumping rate $N_n^{max} = \sqrt{\frac{W_{01}}{2\Gamma^{sp}} + \frac{1}{4}} - \frac{1}{4}$.

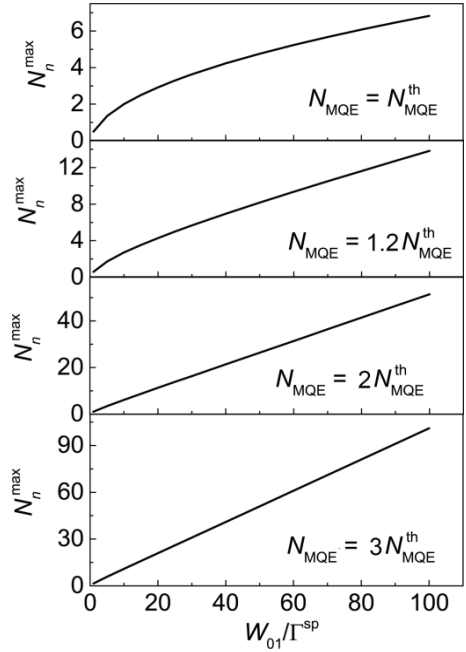

**Fig. 3**. Saturated number of photons $N_n^{max}$ as a function of the normalized pumping rate $W_{01}/\Gamma^{sp}$. The number of MQEs $N_{MQE}$ varies from $N_{MQE}^{th}$ to $3N_{MQE}^{th}$.

## IV. COMPETITIONS BETWEEN MAGNETIC AND ELECTRIC INTERACTIONS

We shall now investigate the coupling strengths of a magnetic interaction and an electric interaction associated with a cavity mode. This is an important factor that determines if a magnetic laser action can exceed an electric one. Note that Eq. (6) also holds for the electric case while the $2\pi g^2$ term should be replaced by $2\pi g_e^2$ to represent the electric interaction. The Purcell factor of an EQE can be expressed as $\Gamma_e^{sp}/\Gamma_0^e$, where $\Gamma_e^{sp}$ and $\Gamma_0^e$ are the spontaneous decay rate of an EQE in a cavity and vacuum, respectively. The ratio of the vacuum decay rate of an EQE to that of a MQE $\Gamma_0^e/\Gamma_0$ can reach several orders of magnitude for a common molecular, while it can be much smaller for a rare-earth ion [5,6]. We assume resonant couplings for both electric and magnetic interactions. Numerical calculations show that, the $\frac{\Gamma^{sp}/\Gamma_0}{\Gamma_e^{sp}/\Gamma_0^e}$ is ~ $10^1$ for a WGM of $n_r = 3.5$ (Fig. 4). This means that $\Gamma_0^e/\Gamma_0$ should be smaller than ~$10^1$ to make the magnetic interaction stronger than the electric one. This can be realistic for rare-earth ions. If only the magnetic interaction is a resonant coupling, the detuning of the nonresonant electric interaction is $\omega - \omega_n = f\gamma_n$. Based on Eq. (2), the nonresonant decay rate of an emitter is $1/(f^2+1)$ times of the resonant one. This will make the ratio $\frac{\Gamma^{sp}/\Gamma_0}{\Gamma_e^{sp}/\Gamma_0^e}$ become relatively $(f^2+1)$ times larger.

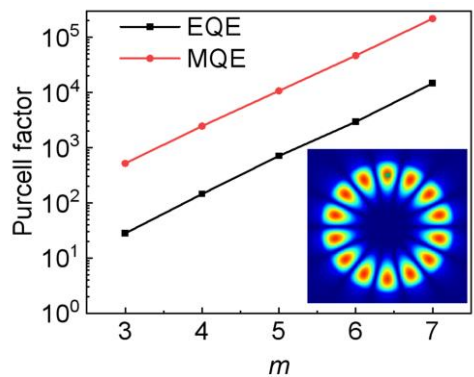

**Fig. 4**. Magnetic Purcell factors for different modes (red dots). Electric Purcell factors with an

EQE as the excitation are also shown (black dots). The disk is the same as that in Fig. 1(b). For each case, an emitter is placed at the maximal field of a mode. The inset shows magnetic field distribution at the center cross section of the disk ($m = 7$). A MQE denoted by the green point is located at a maximal field point of the $m = 7$ WGM mode.

One efficient way to further enlarge the emission ratio $\frac{\Gamma^{sp}/\Gamma_0}{\Gamma_e^{sp}/\Gamma_0^e}$ can be considered by putting an emitter inside a less symmetrical cavity. Here, we also assume resonant couplings for both electric and magnetic interactions for simplicity. Figure 5(a) shows a Si block cavity with an emitter at its center. The length, width and height are 1500, 1050 and 1050 nm, respectively. Here, the size of the cavity is also chosen to match that $n_r$ is around $n_r = 3.5$. There is a high-order magnetic mode around $\lambda = 1375$ nm which is around the size of the cavity (Figs. 5(b)-5(d)). This mode can be efficiently excited by a $y$-polarized MQE with a magnetic Purcell factor of $\Gamma^{sp}/\Gamma_0 \approx 1200$. The $Q$ factor is about $1.5 \times 10^3$. On the other hand, if an EQE is placed at the same point. The $\Gamma_e^{sp}/\Gamma_0^e$ for an EQE polarized in $x$, $y$ and $z$ are only 2.5, 0.3 and 0.3, respectively. We take an average value of $\Gamma_e^{sp}/\Gamma_0^e \approx 1$ for an EQE. The ratio $\frac{\Gamma^{sp}/\Gamma_0}{\Gamma_e^{sp}/\Gamma_0^e}$ can reach $\sim 10^3$. This means that the magnetic interaction can exceed the electric one if $\Gamma_0^e/\Gamma_0$ of an emitter is smaller than $\sim 10^3$. $\frac{\Gamma^{sp}/\Gamma_0}{\Gamma_e^{sp}/\Gamma_0^e}$ increases exponentially with $n_r$ and reaches $\sim 10^5$ around $n_r = 5$ (Fig. 5(e)). Based on Eq. (6), the $N_{MQE}^{th}$ for the above magnetic mode with $n_r = 3.5$ is $N_{MQE}^{th} \sim 10^6$. This number is achievable in such a system. The $Q$ factor of the mode increases almost exponentially with $n_r$. Thus, the $N_{MQE}^{th}$ decreases almost exponentially with $n_r$ (Fig. 5(f)). It is also relatively beneficial to enlarge the cavity size, and the discussion is the same as that in a WGM cavity (Fig. 2).

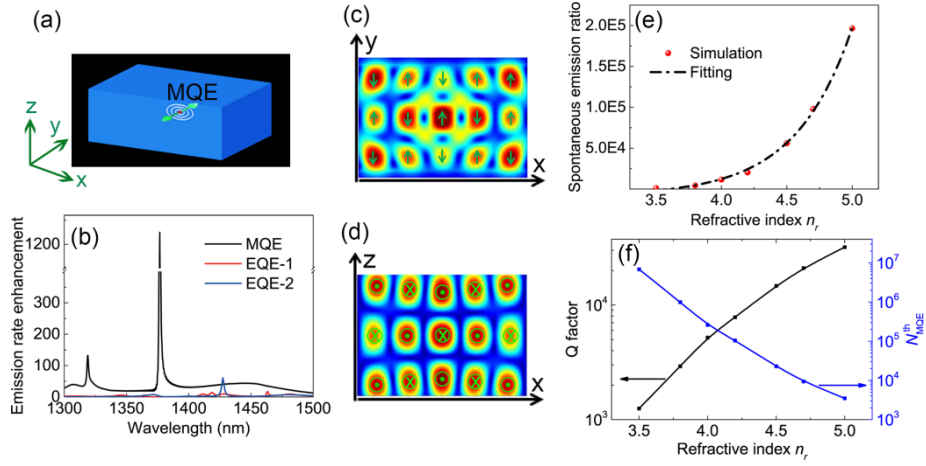

**Fig. 5**. (a) Schematic of a dielectric block excited by a MQE. The origin of the coordinate system is placed at the block center. (b) Spontaneous emission rate enhancement of a MQE ($\Gamma^{sp}/\Gamma_0$) and an EQE ($\Gamma_e^{sp}/\Gamma_0^e$). The polarization of the MQE is along *y*-axis. The polarization of the EQE is along *x*-axis (EQE-1) or *z*-axis (EQE-2). The emitter is located at the block center in each case. $n_r$ = 3.5. (c,d) Magnetic field distribution on the *x-y* (c) and *x-z* (d) plane of the MQE-excited block at $\lambda$= 1375 nm. The arrows denote the main feature of the magnetic field directions. (e) Emission ratio $\frac{\Gamma^{sp}/\Gamma_0}{\Gamma_e^{sp}/\Gamma_0^e}$ as a function of the refractive index $n_r$. (f) *Q*-factor and $N_{MQE}^{th}$ as a function of $n_r$. The size of the block in (e) and (f) is the same as that in (b).

## V. CONCLUSION

In conclusion, we have theoretically proposed that a magnetic laser can be obtained through the stimulated emissions of MQEs in a subwavelength dielectric cavity. The quantum treatment of such a hybrid system is carried out by considering the interactions of quantized magnetic field and two-level MQEs. The magnetic laser can be achieved in a subwavelength cavity based on the facts that the cavity can host high-*Q* electromagnetic resonances with significant magnetic near field responses.

The saturated number of photons $N_n^{max}$ shows a linear relation with the pumping rate when the number of MQEs is more than several times larger than its threshold value. The competition between the electric interaction and magnetic one in terms of their decay rate enhancements is also considered. In a wavelength-scale block cavity, their Purcell factor ratio can reach more than ~$10^3$ large ($n_r$ = 3.5) due to the location dependent emission properties. The widely developed fabrications of combined systems of dielectric structures and rare-earth ions may provide technical support for realizing our proposed magnetic laser in experiments [47-52]. It should be noted that to keep the couplings between MQEs and a cavity efficient enough in experiments, the MQEs should be placed carefully to match the near field distributions of a cavity mode. Our results will enrich the laser field and could find important applications in enhanced magnetic light-matter interactions.

**ACKNOWLEDGMENTS**

This paper was supported by the National Natural Science Foundation of China (No. 11704416), the Hunan Provincial Natural Science Foundation of China (No. 2021JJ20076).

**APPENDIX A: THE INTERACTION HAMILTONIAN BETWEEN A MQE AND THE QUANTIZED MAGNETIC FIELD OF A CAVITY**

The magnetic field of a high-$Q$ electromagnetic mode in a cavity can be expressed as

$$\vec{H}(r,t) = a\vec{Q}(r)\cos(\omega_n t + \varphi(r))$$

$$= \frac{a}{2}\vec{Q}(r)e^{-i(\omega_n t+\varphi(r))} + \frac{a}{2}\vec{Q}(r)e^{i(\omega_n t+\varphi(r))} \tag{A1}$$

where $\varphi(r)$ represents the spatial phase of the magnetic field, $\omega_n$ is the frequency of the photon, and $a$ is the maximal magnitude of the magnetic field ($H_{max}$), namely, $H_{max} = a$. $\vec{Q}(r)$ is a real function of $r$. It represents the normalized magnitude of the magnetic at location $r$, and the maximal value of $\vec{Q}(r)$ is 1. Thus, the combination of $\vec{Q}(r)$ and $a$ (namely, $a\vec{Q}(r)$) means the magnitude of magnetic field at location $r$ (denoted by $\vec{H}(r)$). Thus, we also have $\vec{Q}(r) = \vec{H}(r) / H_{max} = \vec{B}(r) / B_{max}$, where $\vec{B}(r) = \mu_0 \vec{H}(r)$. The time averaged energy of a magnetic mode can be expressed in terms of magnetic field as

$$U_m = \frac{1}{2}\int \mu_0 |\vec{H}(r)|^2 d^3r$$

$$= c^2 a^2, \tag{A2}$$

where $c^2 = \frac{1}{2}\int \mu_0 |\vec{Q}(r)|^2 d^3r$. The quantized Hamiltonian becomes the harmonic oscillator form [37,41,42] provided that $a = \frac{\sqrt{\hbar\omega_n}}{c}\hat{a}$ and $a^* = \frac{\sqrt{\hbar\omega_n}}{c}\hat{a}^+$. Here, $\hat{a}^+$ and $\hat{a}$ are the creation and annihilation operators of a photon, respectively. The quantized magnetic field as a function of position and time can be written as [37,41,42]

$$\vec{H}(r,t) = \frac{\sqrt{\hbar\omega_n}}{2c}\vec{Q}(r)\hat{a}e^{-i(\omega_n t+\varphi(r))} + \frac{\sqrt{\hbar\omega_n}}{2c}\vec{Q}(r)\hat{a}^+ e^{i(\omega_n t+\varphi(r))}. \tag{A3}$$

The interaction Hamiltonian is $H_{int} = -\vec{M}\cdot\vec{B}$, where $\vec{M}$ is the magnetic dipole moment of a MQE. With the second quantization and rotating wave approximation [42], $H_{int}$ can be expressed as

$$H_{int} = -\mu_0 \frac{\sqrt{\hbar\omega_n}}{c}\vec{M}_{10}\cdot\vec{Q}(r)(\hat{a}\hat{\sigma}_{10}e^{-i(\omega_n t+\varphi(r))} + \hat{a}^+\hat{\sigma}_{01}e^{i(\omega_n t+\varphi(r))})$$

$$= -\hbar g(\hat{a}\,\hat{\sigma}_{10}e^{-i(\omega_n t+\varphi(\vec{r}))} + \hat{a}^+\hat{\sigma}_{01}e^{i(\omega_n t+\varphi(\vec{r}))}), \tag{A4}$$

where $\vec{M}_{10}$ is the matrix element of the MD transition, $\hat{\sigma}_{10} = |1\rangle\langle 0|$, and $\hat{\sigma}_{01} = |0\rangle\langle 1|$ are transition operators of a MQE, and $g$ is the coupling strength

$$g = \mu_0 \frac{\sqrt{\hbar\omega_n}}{2c} \vec{M}_{10} \cdot \vec{Q}(r)$$

$$= \mu_0 \frac{\sqrt{\hbar\omega_n}}{2c} \vec{M}_{10} \cdot \frac{\vec{H}(r)}{H_{max}}$$

$$= \mu_0 \frac{\sqrt{\hbar\omega_n}}{2c} \vec{M}_{10} \cdot \frac{\vec{B}(r)}{B_{max}}, \tag{A5}$$

The coupling strength $g$ can also be written in terms of the mode volume of a magnetic mode. The mode volume of a magnetic mode can be expressed as

$$V_m = \frac{\int \mu_0 |\vec{H}(r)|^2 d^3r}{\mu_0 H_{max}^2}$$

$$= \frac{2c^2}{\mu_0}, \tag{A6}$$

Based on Eqs. (A5) and (A6), the coupling strength $g$ can also be written as

$$g = \sqrt{\frac{\mu_0 \omega_n}{2\hbar V_m}} \vec{M}_{10} \cdot \frac{\vec{B}(r)}{B_{max}}. \tag{A7}$$

**APPENDIX B: STIMULATED EMISSION AND ABSORPTION RATE**

Under Fermi's golden rule, the total emission rate into photons from a MQE can be expressed as

$$\Gamma' = \frac{2\pi}{\hbar^2} |\langle N_n + 1, 0|H_{int}|N_n, 1\rangle|^2 \delta(\omega_{10} - \omega_n), \tag{B1}$$

where $\omega_{10}$ is the frequency of MQE transition. $|N_n, 1\rangle$ represents a state where there are $N_n$ photons and the MQE is on the excited state. Similarly, $|N_n + 1, 0\rangle$ represents a state where there are $N_n + 1$ photons and the MQE is on the ground state. Based on Eq. (A4), the magnitude matrix element for the emission of a photon can be written as

$$|\langle N_n + 1, 0|H_{int}|N_n, 1\rangle| = \hbar g\sqrt{N_n + 1}. \tag{B2}$$

Considering certain spectral widths of both MQE transitions and photons [37], Eq.

(B1) can be expressed as

$$\Gamma' = 2\pi g^2 (N_n + 1) \int F(\omega) \frac{\gamma_n^2}{(\omega - \omega_n)^2 + \gamma_n^2} d\omega, \tag{B3}$$

where $\int F(\omega) \frac{\gamma_n^2}{(\omega - \omega_n)^2 + \gamma_n^2} d\omega$ is the spectral overlap factor. $F(\omega)$ is the normalized-to-1 spectrum of MD transitions, and $\gamma_n$ is the relaxation rate of the photon. In the $N_n + 1$ term of Eq. (B3), the $N_n$ represents the contribution from stimulated emission $\Gamma^{st}$ while the number 1 corresponds to the spontaneous emission $\Gamma^{sp}$. Similar to Eq. (B2), the magnitude matrix element for the absorption of a photon can be written as

$$|\langle N_n - 1, 1|H_{int}|N_n, 0\rangle| = \hbar g\sqrt{N_n}. \tag{B4}$$

Thus, one can obtains the stimulated absorption rate $\Gamma^{sa}$, and it is equal to $\Gamma^{st}$.

## APPENDIX C: THE RATE EQUATION FOR THE POPULATION OF MQES

The population density $\rho_1$ of MQEs at a given location satisfies

$$\dot{\rho}_1 = \rho_0 W_{01} + \rho_0 \Gamma^{st} - \rho_1 \Gamma^{st} - \rho_1 \Gamma^{sp}, \tag{C1}$$

where $W_{01}$ is the pumping rate. Similarly, $\rho_0$ satisfies $\dot{\rho}_0 = -\dot{\rho}_1$. We can define $\rho_{tot} = \rho_1 + \rho_0$, and we also have $\rho_{eff} = \rho_1 - \rho_0$. Thus, $\dot{\rho}_{eff} = 2\dot{\rho}_1$. Then, $\rho_1$ and $\rho_0$ can be expressed as $2\rho_1 = \rho_{tot} + \rho_{eff}$ and $2\rho_0 = \rho_{tot} - \rho_{eff}$, respectively. Eq. (C1) can be rewritten in terms of $\rho_{eff}$ and $\rho_{tot}$ as

$$\dot{\rho}_{eff} = (W_{01} - \Gamma^{sp})\rho_{tot} - (W_{01} + \Gamma^{sp} + 2\Gamma^{st})\rho_{eff}. \tag{C2}$$

Considering the integration over the whole region, one obtains Eq. (4).

## APPENDIX D: THE STEADY-STATE SOLUTION FOR THE SATURATED NUMBER OF PHOTONS

The $N_n^{max}$ can be obtained by solving Eqs. 3 and 4 under the steady-state conditions, namely, $\dot{\rho}_{eff} = 0$ and $\dot{N}_n = 0$. There are two solutions of $N_n^{max}$. The nagtive one is omitted. The positive one is

$$N_n^{max} = \frac{1}{4} - \frac{W_{01}}{4\Gamma^{sp}} + \frac{W_{01}N_{MQE}}{4\gamma_n} + \left[\frac{1}{16} + \frac{W_{01}}{8\Gamma^{sp}} + \frac{3W_{01}N_{MQE}}{8\gamma_n} + \left(\frac{W_{01}}{4\Gamma^{sp}} - \frac{W_{01}N_{MQE}}{4\gamma_n}\right)^2\right]^{1/2}.$$

(D1)

When $N_{MQE}$ is several times larger than $N_{MQE}^{th}$ ($N_{MQE}^{th} = \frac{\gamma_n}{\Gamma^{sp}}$), the expression $\left[\frac{1}{16} + \frac{W_{01}}{8\Gamma^{sp}} + \frac{3W_{01}N_{MQE}}{8\gamma_n} + \left(\frac{W_{01}}{4\Gamma^{sp}} - \frac{W_{01}N_{MQE}}{4\gamma_n}\right)^2\right]$ in Eq. (D1) is dominated by the $\left(\frac{W_{01}}{4\Gamma^{sp}} - \frac{W_{01}N_{MQE}}{4\gamma_n}\right)^2$ term. Thus, the $N_n^{max}$ becomes

$$N_n^{max} \approx \frac{1}{4} + \frac{(N_{MQE} - N_{MQE}^{th})W_{01}}{2\Gamma^{sp}}.$$

(D2)

When $N_{MQE}$ is equal to $N_{MQE}^{th}$, the $\frac{W_{01}}{4\Gamma^{sp}} - \frac{W_{01}N_{MQE}}{4\gamma_n}$ term becomes 0. Thus, the $N_n^{max}$ is

$$N_n^{max} = \sqrt{\frac{W_{01}}{2\Gamma^{sp}} + \frac{1}{4}} - \frac{1}{4},$$

(D3)

namely, $N_n^{max} \propto \sqrt{\frac{W_{01}}{\Gamma^{sp}}}$.